\documentclass[%
 reprint,
 amsmath,amssymb,
 aps,
]{revtex4-2}
\usepackage{xcolor}
\usepackage{graphicx}
\usepackage{dcolumn}
\usepackage{bm}
\usepackage{xcolor}
\usepackage{placeins}
\usepackage[utf8]{inputenc}
\usepackage[T1]{fontenc}
\usepackage{mathptmx}
\usepackage{etoolbox}
\usepackage[siunitx]{circuitikz}
\usepackage{adjustbox}
\usepackage[font = bf]{subcaption}
\usepackage{caption}
\usepackage{comment}
\usepackage{tikz}

\usetikzlibrary{decorations.pathmorphing}
\tikzset{%
lray/.style={decorate, decoration={
 snake, amplitude=2pt,pre length=1pt,post length=2pt, segment length=5pt,},
 -Triangle,
 }}
\tikzset{%
 llray/.style={decorate, decoration={
 snake, amplitude=8pt,pre length=4pt,post length=8pt, segment length=5pt,},
 -Triangle,
 }}
 \tikzset{%
 lllray/.style={decorate, decoration={
 snake, amplitude=15pt,pre length=4pt,post length=6pt, segment length=7pt,},
 -Triangle,
 }}
 \newcommand{\micron}{$\mu$m}
 \newcommand{\hot}{\mathrm{H}}
 \newcommand{\cold}{\mathrm{C}}
 \newcommand{\term}{\mathrm{T}}

\begin{document}

\preprint{APS/123-QED}

\title{A  Near Quantum Limited Sub-GHz TiN Kinetic Inductance Traveling Wave Parametric Amplifier Operating in a Frequency Translating Mode }

\author{Farzad Faramarzi}
 \email{farzad.faramarzi@jpl.nasa.gov.}

\author{Sasha Sypkens } 

\author{Ryan Stephenson}

\author{Byeong H. Eom}

\author{Henry LeDuc}

\author{Saptarshi Chaudhuri}

\author{Peter Day}%
    \email{peter.k.day@jpl.nasa.gov}

\affiliation{Jet Propulsion Laboratory, California Institute of Technology, Pasadena, CA 91101,USA}%

\affiliation{California Institute of Technology, 1200 California Blvd., Pasadena, 91125, CA, USA}

\affiliation{ Department of Physics, Princeton University, Princeton, NJ 08544, USA}

\date{\today}

\begin{abstract}
    We present the design and experimental characterization of a kinetic-inductance traveling-wave parametric amplifier (KI-TWPA) for sub-GHz frequencies. KI-TWPAs amplify signals through mixing processes supported by the nonlinear kinetic inductance of a superconducting transmission line. The device described here utilizes a compactly meandered TiN microstrip transmission line to achieve the length needed to amplify sub-GHz signals.  It is operated in a frequency translating mode where the amplified signal tone is terminated at the output of the amplifier, and the idler tone at approximately 2.5~GHz is brought out of the cryostat.  By varying the pump frequency, a gain of up to 22 dB was achieved in a tunable range from about 450 to 850~MHz. Use of TiN as the nonlinear element allows for a reduction of the required pump power by roughly an order of magnitude relative to NbTiN, which has been used for previous KI-TWPA implementations. This amplifier has the potential to enable high-sensitivity and high-speed measurements in a wide range of applications, such as quantum computing, astrophysics, and dark matter detection. \footnote{© 2024. All rights reserved}

\end{abstract}

\maketitle

\section{Introduction}

Kinetic inductance traveling-wave parametric amplifiers (KI-TWPAs)\cite{eom_day_leduc_zmuidzinas_2012} with near quantum-limited noise performance and wideband gain have been developed recently and used for the readout of low-temperature detectors and qubits~\cite{zorbist,mumux,malnou,ranzani2018}.  The application of KI-TWPAs to direct dark matter searches is also gaining attention~\cite{karthik,ramanathan2024}. The majority of these amplifiers have been designed for the 4-8~GHz band. However, a few technologies and experiments in quantum information science, astronomy, and fundamental physics can benefit from a near-quantum-noise-limited amplifier with gain at sub-GHz frequencies. For example, single-shot readout of spin-qubits operating below 1~GHz may be possible using a quantum-limited sub-GHz amplifier~\cite{Qdot1,Qdot2,Qdot3,Spinqubit1}.
Reducing the readout noise of nano-calorimeters and nano-bolometers would allow them to be sensitive to single photons in the microwave regime~\cite{nanobolojj,pekola,nanocalo,zepto,workheat,opensys}. Circuit quantum cavity electrodynamics with micro-mechanical hybrid systems can also benefit from sub-GHz quantum limited amplifiers~\cite{Pirkkalainen,Palomaki}. Far- and mid-IR MKID detectors operating in the single photon counting regime\cite{PeterPRIMA} could be improved with lower noise readout amplifiers operating around 1~GHz. 

Direct searches for ultralight dark matter, such as axions and hidden photons, can also benefit from quantum-limited sub-GHz amplifiers with wide bandwidth. For example, the Princeton Axion Search ~\cite{chaudhuri2024introducing} aims to probe QCD axion dark matter in the 0.8-2 $\mu$eV mass range, corresponding to frequencies 200-500 MHz; a quantum-limited amplifier would enable the search to achieve the baseline noise performance. Moreover, the use of traveling-wave amplifiers with wide bandwidth, as opposed to narrowband resonant amplifiers, may allow for simplified receiver design ~\cite{bartram2023dark} and largely eliminates the need for frequent tuning of the amplifier band in a wideband scan.

The best High Electron Mobility Transistor (HEMT) amplifiers made of InP claim noise temperatures close to 1.5 K ~\cite{cha2023optimization,LNF}, corresponding to $\sim 31h\nu$ per second per Hz (``photons'') at 1~GHz. As these amplifiers dissipate on the order of $\sim 1$~mW, they must be located at a higher temperature stage of a sub-kelvin cryostat, typically following an isolator, incurring losses that increase the effective noise. Sub-GHz Josephson parametric amplifiers (JPAs) with near-quantum limited performance have been developed recently. However, due to their resonant nature, they have low bandwidths and very low saturation power~\cite{subghzjj1,simbi}.

\begin{figure*}
\centering
     \begin{subfigure}[b]{\textwidth}
     \caption{}
         \centering
         \includegraphics[width=0.95\textwidth]{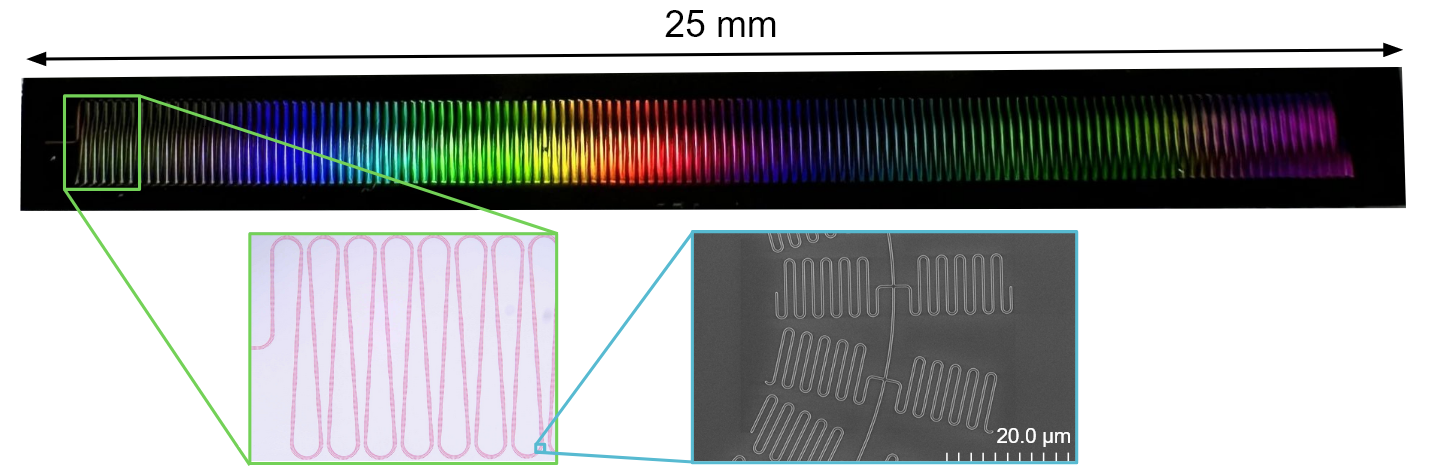}
         \label{fig:a}
     \end{subfigure}
     
     \begin{subfigure}[b]{0.95\columnwidth}
     \caption{}
         \centering
         \includegraphics[width=\columnwidth]{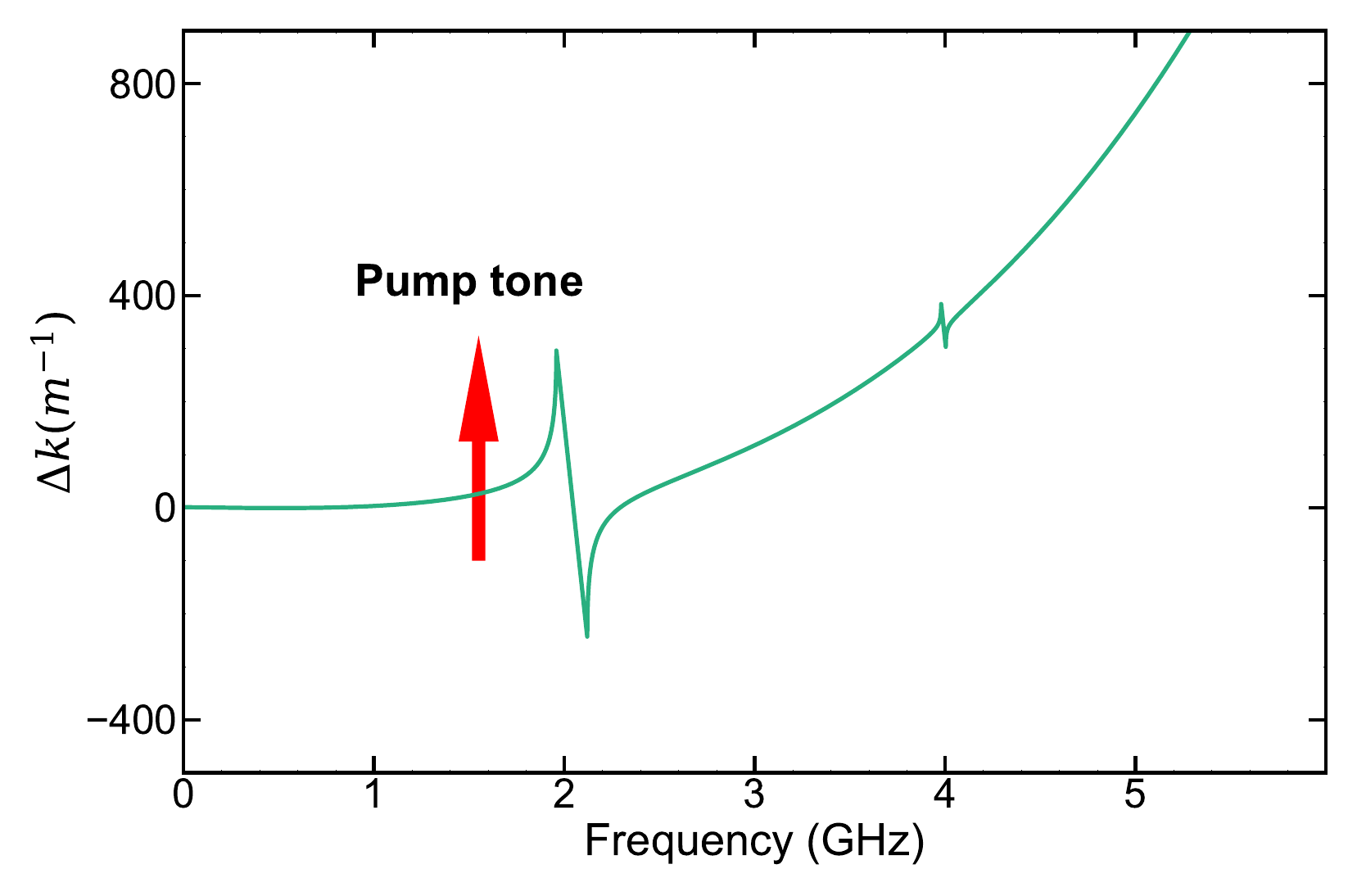}
         \label{fig:b}
     \end{subfigure}
     \begin{subfigure}[b]{0.95\columnwidth}
     \caption{}
         \centering
         \includegraphics[width=\columnwidth]{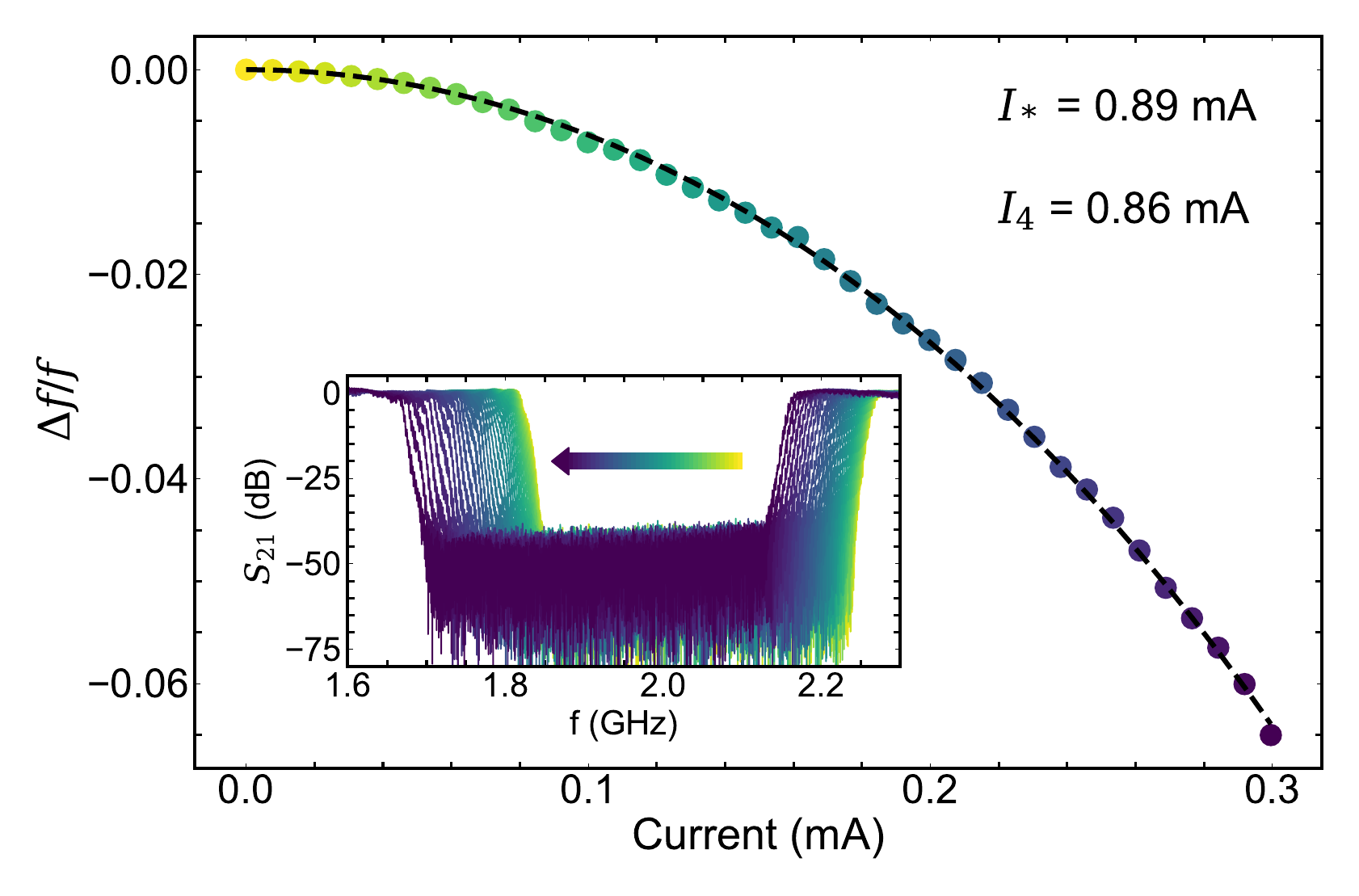}
         \label{fig:b}
     \end{subfigure}
        \caption{\label{disper:gain} a) Picture of the sub-GHz KI-TWPA chip. The total length of the device is 450~mm; however, the chip's length is 25~mm, including the meandered KI-TWPA transmission line and the bond pads. The inset to the left is the zoomed-in image showing the meandered transmission line geometry. The inset to the right is the SEM micrograph of the capacitive fingers and the microstrip center line with a width of 320~nm. The capacitive fingers are meandered for compactness. b) Plot of the calculated $\Delta k$, the deviation of dispersion of the KI-TWPA from a dispersionless line. 
        c) Plot of the measured fractional frequency shift of the center of the bandgap as a function of DC bias. From a fit to equation~\ref{eq:bndgpshift} the extracted characteristic currents are $I_\ast = 0.89 \mathrm{mA}$ and $I_4 = 0.86 \mathrm{mA}$. The inset shows the measured $S_{21}$ of the bandgap as a function of applied current.}
        \label{fig:device}
\end{figure*}

This article presents results on a wide bandwidth sub-GHz KI-TWPA with sensitivity close to the quantum limit.  The design of the amplifier draws on our earlier work\cite{shibo,4G8GFF}, but introduces two new innovations.  The first departure from previous work is the use of a TiN microstrip line in place of NbTiN as the nonlinear transmission line, which significantly reduces the required pump power.  Reduced pump power facilitates thermal control of the experimental setup at the low temperatures required to avoid thermal noise.  For example, at 500~MHz, 1 photon corresponds to $hf / k_\mathrm{B} = 24$~mK.  KI-TWPA designs based on 4-wave mixing (4WM) typically needed a pump power of $\sim 10\,\mu$W.  While the KI-TWPA itself dissipates only a fraction of the pump power, it is harder to control dissipation in the surrounding circuitry, including microwave isolator.  The TiN device described here requires only $\sim 0.3\,\mu$W, which causes minimal heating at the base temperature of a dilution refrigerator.

The second innovation enabling a practical and tunable sub-GHz amplifier is the operation of the KI-TWPA as both amplifier and frequency up-converter.  We accomplish this by designing the device to produce an idler tone at 3 to 6 times the signal frequency.  While the signal band center can be tuned over about an octave in frequency, from 450 to 850~MHz, the idler band center remains relatively fixed near 2.5~GHz.  Because the idler is an exact replica of the signal down to the level of quantum fluctuations, we bring only the idler through the following amplification stages without loss of information.  This frequency translating gain mode has important advantages.  First, sub-GHz cryogenic isolators with wide bandwidth do not exist.  Our circuit can use a relatively narrow band isolator at the idler frequency. Second, the large frequency conversion factor confers an additional power gain factor of the ratio of the idler to signal frequencies $\omega_i / \omega_s$, because the idler photons are produced on a one-to-one basis with the signal photons.  Third, the circuit is more tolerant of thermal noise, as the added noise of the TWPA corresponds to the thermal noise at the idler frequency.  Thermal noise may be significant at the sub-GHz signal frequencies for achievable temperatures while the higher frequency idler modes remain in the vacuum state.

\section{Device Design}

Similar to KI-TWPAs produced previously in our laboratory, the device reported here is based on a nonlinear transmission line (NTRL) structure with an inverted microstrip line geometry \cite{shibo, nikita, 4G8GFF}.  The microstrip conductor layer is patterned from a 50~nm thick TiN film.  The dielectric of the microstrip is a 100~nm thick amorphous Si layer, and the sky plane is 200~nm thick niobium.  The width of the microstrip line is 320~nm.  Every 15~\micron\, a pair of microstrip open stubs of the same width and average length 120~\micron\, are connected to the central microstrip line.  The NTRL is a slow-wave structure with propagation velocity $\sim 0.0064c $ and is tightly meandered, as shown in fig.~\ref{fig:device}a., across a 25~mm long chip.  The 45~cm total length of the NTRL is several times longer than KI-TWPAs designed for >~4~GHz in order to roughly maintain the number of wavelengths ($\sim$300 pump wavelengths) for sufficient gain at the sub-GHz signal frequencies.   

The length of the stub sections is also several times greater than that of the higher frequency devices.  The stub length is chosen to add sufficient dispersion to the NTRL to prevent phase matching of unwanted parametric processes such as third harmonic generation.  A plot of the deviation of the propagation constant versus frequency $\Delta k(\omega)$ with respect to a dispersionless ($k \propto \omega$) behavior is shown in fig.~\ref{fig:device}b.  The upward curvature past about 3~GHz is due to the length of the stubs.  Having achieved the desired degree of dispersion, the spacing of the stubs was determined so that their added capacitance per unit length reduced the characteristic impedance of the NTRL structure to approximately 50~$\Omega$.  The feature in $\Delta k(\omega)$ near 2~GHz results from a sinusoidal modulation of the stub length with amplitude 20~\micron\, and period of 461~\micron\, that produces a stop band in the transmission (shown in the inset of panel c of the figure.)  



To the lowest order, the NTRL responds quadratically to current via the current dependence of the kinetic inductance 
$\delta L_\mathrm{K}(I)/{L_\mathrm{K}(0)} \approx {I^2}/{I^2_\ast}$,
where \cite{shibo}
\begin{equation}
    I_\ast = wt\kappa_*\frac{\Delta}{\lambda_L}\sqrt{\frac{N_0}{\mu_0 }}.
\label{eqn:Istar}
\end{equation}
$N_0$ is the single-spin density of states at the Fermi level, $\Delta$ is the gap parameter, $\lambda_\mathrm{L}$ is the penetration depth, $\kappa_\ast \approx 1$ and $w$ and $t$ are the microstrip conductor width and thickness.  Changing from NbTiN to TiN reduces $I_\ast$ by approximately the ratio of the transition temperatures $T_\mathrm{c} \propto \Delta$, which are 4.5~K and 12.5~K for TiN and NbTiN, respectively.  The other parameters in eqn.~\ref{eqn:Istar} are similar for the two materials. The pump power is expected to vary roughly as $\Delta^2$.  The value of $I_\ast$ for the TiN NTRL was determined by measuring the shift in the frequency of the stop band center, $f_{\mathrm{gap}}$ versus injected DC current (fig.~\ref{fig:device}.c) \cite{shibo}.  The data were fit to a quartic dependence in the applied current:
\begin{equation}
    \frac{f_{\mathrm{gap}}(I_{\mathrm{dc}}) - f_{\mathrm{gap}}(0)}{f_{\mathrm{gap}}(0)} = -\frac{1}{2} \bigg[ \bigg( \frac{I_{\mathrm{dc}}}{I_\ast} \bigg)^2 + \bigg( \frac{I_{\mathrm{dc}}}{I_4} \bigg)^4 \bigg].
    \label{eq:bndgpshift}
\end{equation}
The fit yields $I_\ast = 0.89$~mA and $I_4 = 0.86$~mA.  The $I_\ast$ value is about 3.6 times less than for an NbTiN microstrip line of the same width, but a thickness of 35~nm, reported in \cite{4G8GFF}.

\begin{figure}
     \centering
     \begin{subfigure}[b]{0.9\columnwidth}
     \caption{}
         \centering
         \includegraphics[width=\columnwidth]{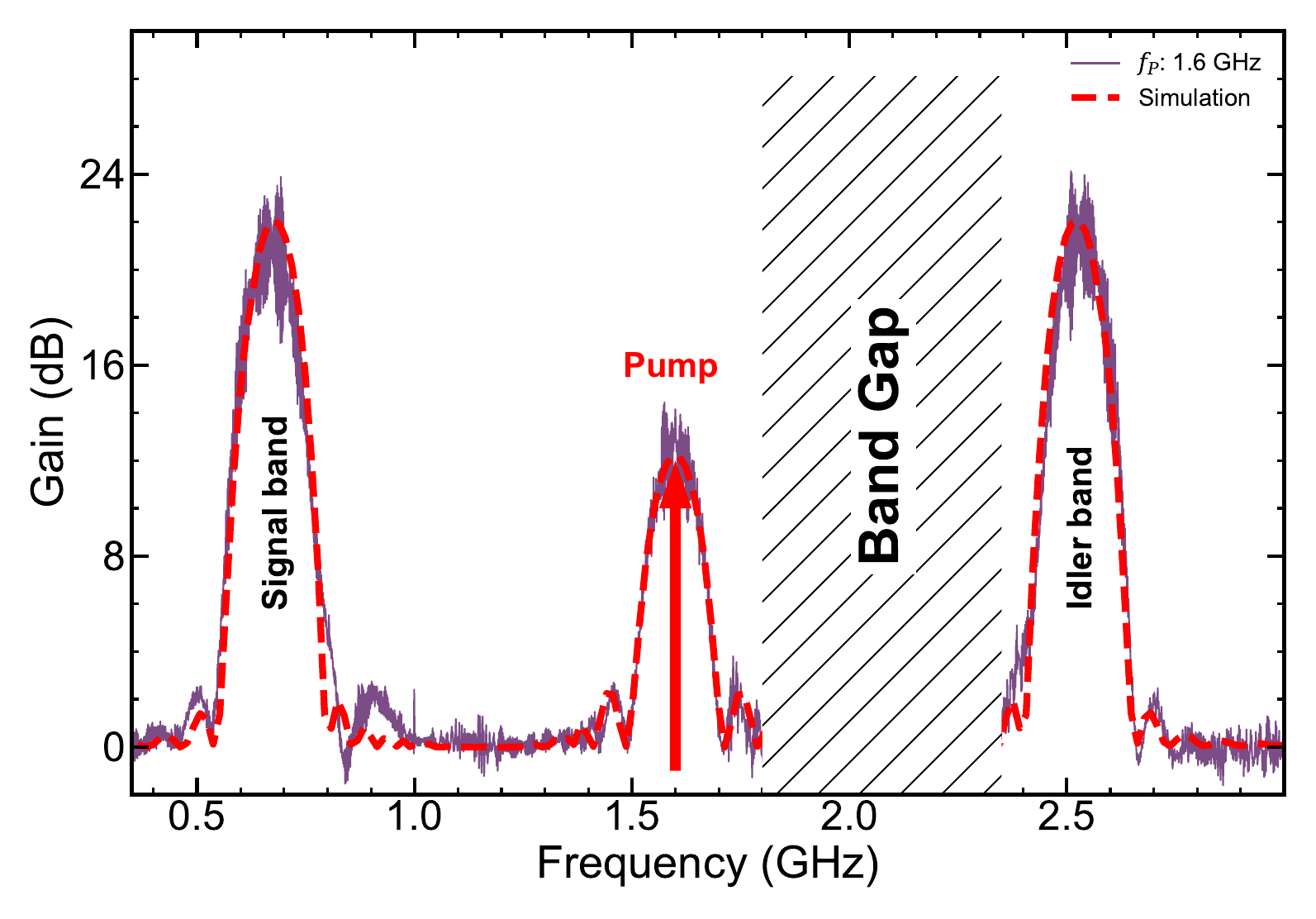}
         \label{fig:a}
     \end{subfigure}
     \begin{subfigure}[b]{0.9\columnwidth}
     \caption{}
         \centering
         \begin{adjustbox}{scale=0.8}
          \begin{circuitikz}[american]
        \ctikzset{resistors/scale=0.5,bipoles/length=2cm}
    \draw
    (-2.5,5) -- (-1,5)
    (0,6)--(0,5.72)
    (0,8) to [twoport,t = DPX 2] (0,6)
    (-1,5) to [twoport,t = DPX 1] (1,5)

    (1.4,8.6)node[color=violet]{ Pump in}
    (1.4,8.1)node[color=violet]{ $\omega_P$}
    (2.8,8.4)node[color=violet, text width=1cm, align = center]{ Pump out} 
    (-2,6.5)node[color=blue]{Signal in}
    (-2,6)node[color=blue]{$\omega_s$}
    (5.8,3.5) node[color=red]{$\omega_i$}
    (5.8,4) node[color=red]{Idler Out}
    (5.8,8) node[color=blue]{Signal Out}

    (1,5) to [amp,t = TWPA] (3,5)
    (3,5) to [twoport,t = DPX 2] (5,5)
    (4,6)--(4,5.72)
    (4,8) to [twoport,t = DPX 1] (4,6)
    (4.6,7) to [R] (6.7, 7) -- (6.7, 6.5) node[tlground]{}
    (5,5)--(6.5,5)
    ;
    \draw [dashed] (6.5,5) to (7,5);
    \draw [dashed] (-2.5,5) to (-3,5);
    \draw[dashed] (0,8) to (0,8.5);
    \draw[dashed] (4,8) to (4,8.5);
    \draw [lllray,color=violet] (1.4,7.8) to (1.4,6.8);
    \draw [llray,color=violet] (2.8,6.8) to (2.8,7.8);
    \draw[lray,color=blue] (-2.5,5.5) to (-1.5,5.5);
    \draw[llray,color=red] (5.2,4.5) to (6.4,4.5);
    \draw[llray,color=blue] (5.2,7.5) to (6.4,7.5);

         \end{circuitikz}
         \end{adjustbox}
         \label{fig:b}
     \end{subfigure}

        \caption{\label{fig:ftg} a) Plot of the gain (pump on / pump off transmission) of the device with an applied -34.8 dBm pump tone at 1.6~GHz, measured using a VNA.  The input and output frequencies are the same for this measurement. The dashed line is the gain calculated by integrating coupled mode equations.\cite{shibo,klimovich2024investigating}  b) The device may also be operated as a frequency translating amplifier using a set of diplexers.  At its input, the signal band is admitted while the device sees a termination over the idler band.  At the output, the idler band is transmitted, and the signal band is terminated.}
        \label{fig:gain}
\end{figure}

\section{Performance}

\begin{figure*}[]
     \centering
     \begin{subfigure}[b]{0.85\columnwidth}
     \caption{}
         \centering
         \includegraphics[width=0.85\columnwidth]{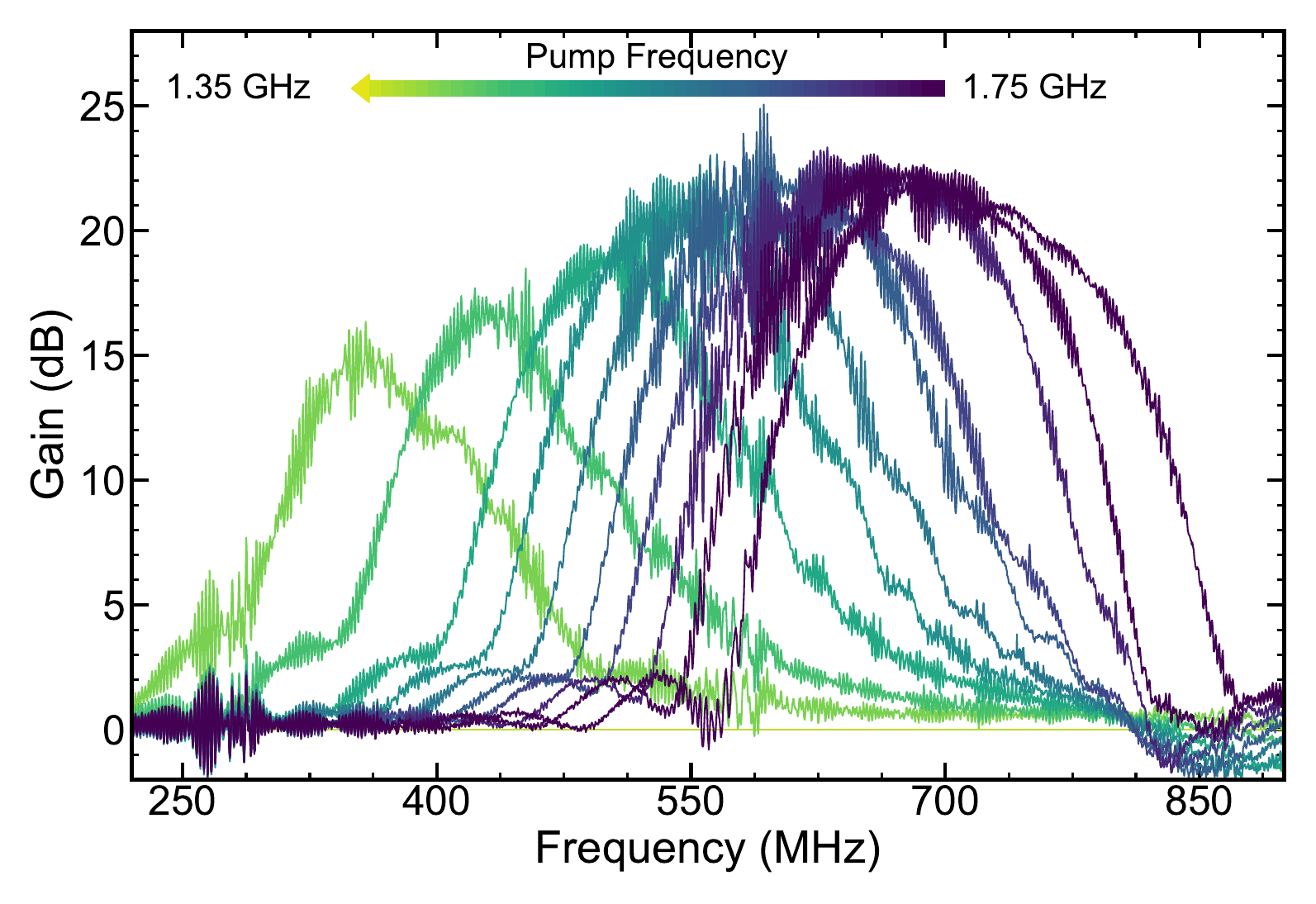}
         \label{fig:a}
     \end{subfigure}
     \begin{subfigure}[b]{0.85\columnwidth}
     \caption{}
         \centering
         \includegraphics[width=0.85\columnwidth]{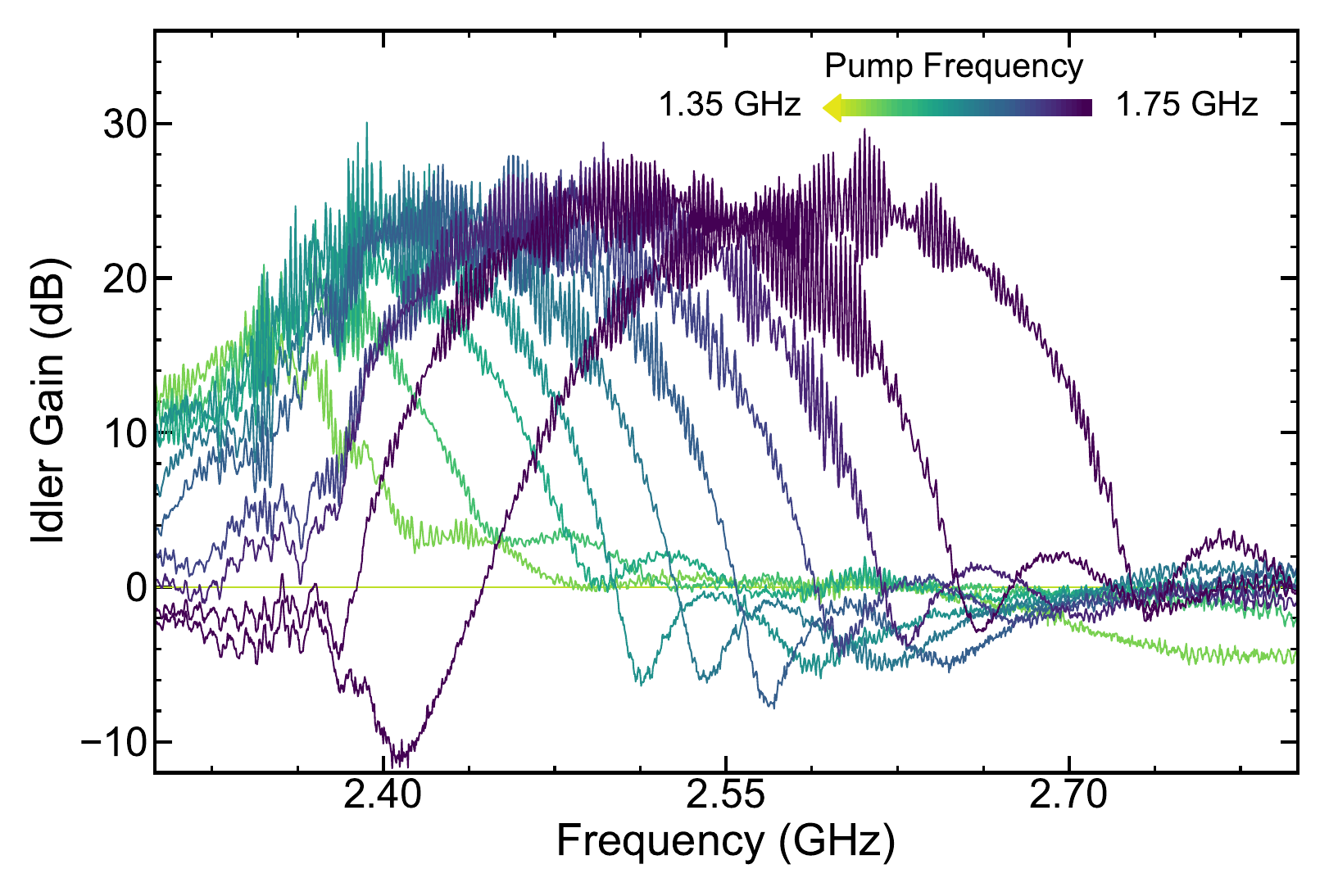}
         \label{fig:b}
     \end{subfigure}
     \begin{subfigure}[b]{0.85\columnwidth}
     \caption{}
         \centering
         \includegraphics[width=0.85\columnwidth]{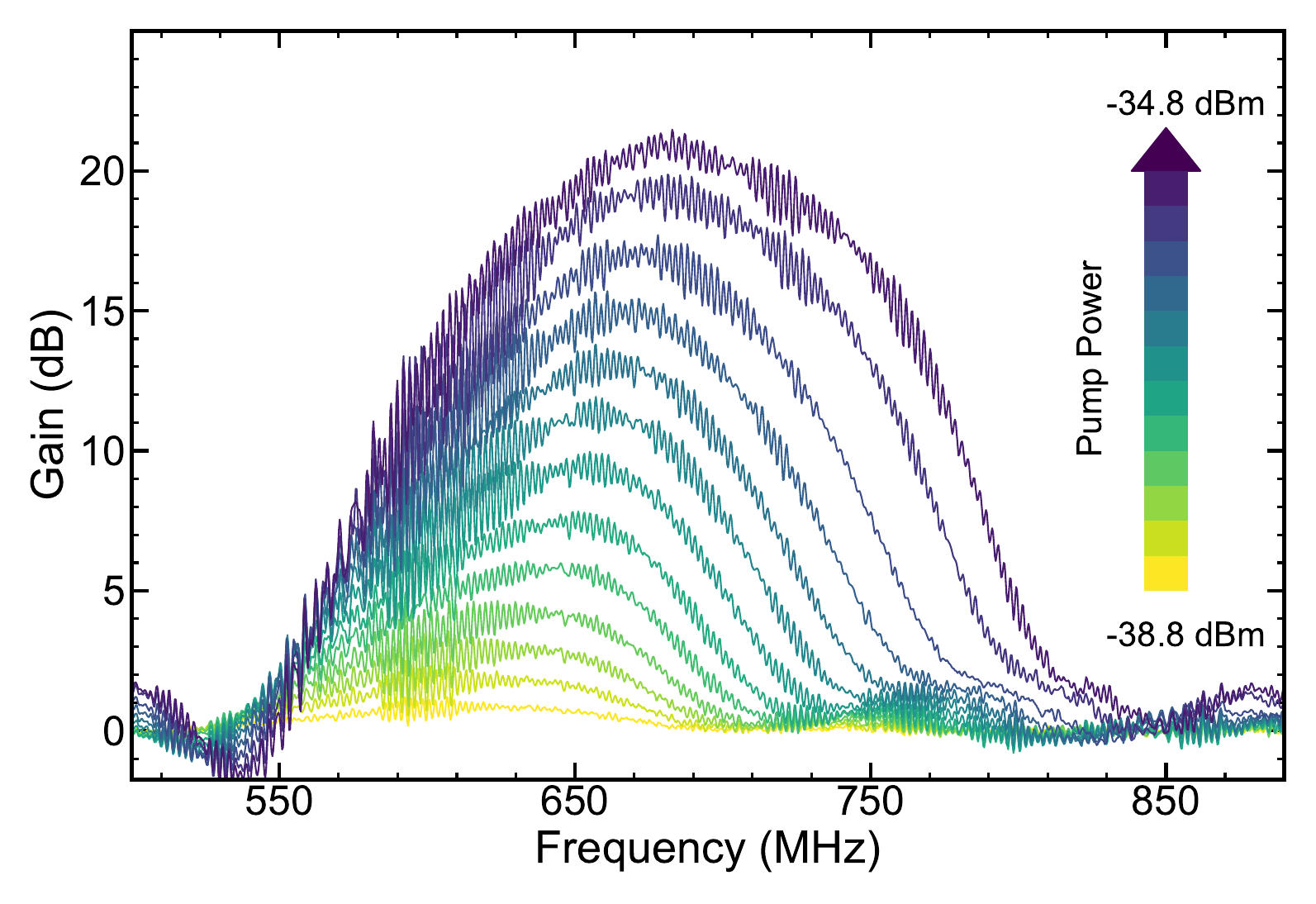}
         \label{figc}
     \end{subfigure}
     \begin{subfigure}[b]{0.85\columnwidth}
     \caption{}
         \centering
         \includegraphics[width=0.85\columnwidth]{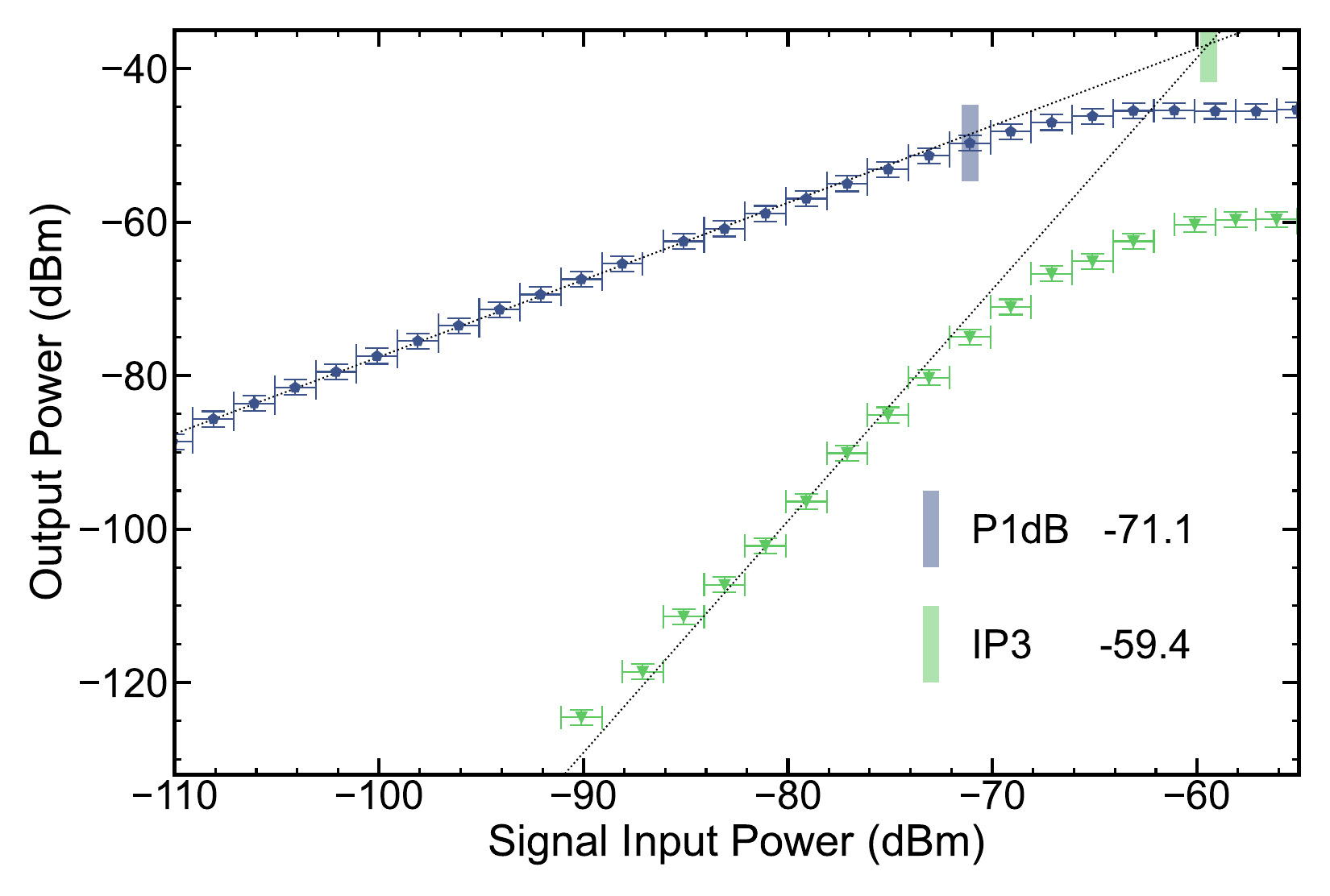}
         \label{fig:d}
     \end{subfigure}
        \caption{\label{fig:gain} Measurements in subfigures a,b, and c were taken at 200~mK with a 20~dB attenuator at the 4~K stage.  A 20~dB directional coupler was used at room temperature to couple the pump and signal tones into the cryostat. a)  Measured signal band gain for a range of pump frequencies and all with a pump power of -34.8~dBm at the device. b)  Corresponding idler band gain.  c) Plot of the measured signal gain at 200~mK as a function of pump power for a pump tone at 1.6~GHz. The maximum gain near 20~dB corresponds to a pump power of 330~nW. At higher pump power the device switches to the normal state. d) A plot of the $\mathrm{P_{1dB}}$ and $\mathrm{IP_3}$ measurements is presented, taken with a pump tone at 1.6 GHz and a gain of $\sim$21.5 dB.}
        \label{fig:gaindetail}
\end{figure*}
Applying a pumping tone below the stop band frequency results in parametric amplification via 4WM for weaker signal tones.  The typical gain shown in fig.~\ref{fig:ftg}.a has three disjointed bands where gain is produced, similar to the device described in \cite{4G8GFF}, but with a wider separation between bands.   The low and high frequency bands reflect the frequencies at which the phase-matching criterion for 4WM is approximately satisfied:
\begin{equation}
    k(\omega_s) + k(\omega_i) - 2 k(\omega_p) \approx - k(\omega_p) \; \frac{I_p^2}{4 I_*^2} \,,
\label{eqn:phasematching}
\end{equation}
where $I_p$ is the pump current, the $k(\omega_{s,i,p})$ are the propagation constants at the signal, idler and pump frequencies, and $2\omega_p = \omega_s + \omega_i$.  The central band reflects non-phase-matched gain around the pump frequency.  The dashed line in the figure shows the gain calculated by integrating the coupled mode equations \cite{shibo,klimovich2024investigating} including third harmonic and upper sideband generation (tones at $3\omega_p$, $2\omega_p+\omega_s$ and $2\omega_p+\omega_i$).  The film penetration depth in the model was adjusted to 625~nm to match the measured stop band frequency.  To match the overall gain, the pump current was set to $I_p/I_\ast = 0.117$, corresponding to 300~nW, which is close to the estimate of the experimental pump power.  

The lower and upper gain bands in the figure are arbitrarily labeled as the ``signal'' and ``idler'' bands.  With the pump slightly below the stop band, input power in either the signal or idler bands stimulates the conversion of pairs of pump photons into correlated signal and idler photons.  As described earlier, it is advantageous to operate the device as a frequency translating amplifier.  In that mode, the power gain for input frequency $\omega_s$ and output frequency $\omega_i$ is \cite{caves1982quantum}
\begin{equation}
    G_{si} = (G_{ss} - 1) \frac{\omega_i}{\omega_s},
    \label{eqn:Gsi}
\end{equation}
where $G_{ss}$ is the gain for both input and output at the signal frequency, and $G_{ss} = G_{ii}$, the idler to idler frequency gain.
This frequency translating operating mode can be easily implemented using the circuit shown in fig.~\ref{fig:ftg}.b.  A pair of diplexers at the input of the KI-TWPA allows for the injection of the pump and provides a cold input load at the idler frequencies.  An output side diplexer filters out the idler frequencies for subsequent amplification while sending the pump and signal band to a termination.

The signal band center can be shifted by almost a factor of two in frequency by adjusting the frequency of the pump tone, as shown in fig.~\ref{fig:gaindetail}.a.  As shown in panel b, the idler band also shifts with pump frequency, but by a smaller amount, because of the greater curvature of the dispersion curve at the idler frequencies (fig.~\ref{fig:device}.b).  This behavior allows for the use of a narrow band isolator for frequency translating gain operation.  

Fig.~\ref{fig:gaindetail}.c shows the response of the gain to changes in pump and signal power.  The maximum gain of 20~dB is reached at a pump power of 330~nW. 
The 1 dB compression point ($\mathrm{P_{1dB}}$) was determined by gradually increasing the input power of a signal tone and measuring the corresponding output power using a spectrum analyzer. For this measurement, an input signal frequency of 675 MHz was selected, while a pump tone at 1.6 GHz with an input power of -35 dBm was applied to achieve a gain of $\sim$ 21.5 dB. The third-order intercept point ($\mathrm{IP_3}$) was measured using two signal tones at 675~MHz and 655~MHz, each with equal input power with the same pump conditions used in the $\mathrm{P_{1dB}}$ measurement. The input powers of the signal tones were gradually increased, and the corresponding output powers were recorded using a spectrum analyzer. Additionally, the output powers of the third-order intermodulation products ($f_1 - f_2$) and ($f_2 - f_1$) were measured to determine the third-order intercept point ($\mathrm{IP_3}$). At a gain of approximately 21.5~dB, the $\mathrm{P_{1dB}}$ was measured to be -71.1~dBm, and the $\mathrm{IP_3}$) was -59.4~dBm. A plot showing the compression point at different gain levels is included in Appendix~C for further reference.

Gain compression for a 4WM TWPA results from pump depletion and sets in when the output power of the signal and idler tones becomes large enough to significantly perturb the pump power.

\section{Noise Measurements}

A Y-factor measurement was used to determine the noise added by the KI-TWPA and the total noise of the measurement system.  A detailed diagram of the frequency translating mode noise measurement circuit is shown in fig.~\ref{fig:LFsetup} in the appendix.  Two 50~$\mathrm{\Omega}$ terminations were used as the noise sources, one at $T_\mathrm{cold} = 10$~mK, thermalized to the mixing chamber (MXC) stage of a dilution refrigerator, and the other at the 3 K stage ($T_\mathrm{hot} = 3.13$~K).  The hot termination is connected to the MXC stage through a superconducting cable with negligible loss over the signal band.  A cryogenic relay switch at the input of the KI-TWPA allowed for switching between the 'hot' and 'cold' sources.  A pair of diplexers at the input of the KI-TWPA were used to limit the noise power from the sources to the signal band of the KI-TWPA, inject the pump tone, and inject a separate test tone over the idler frequency range for measurement of the gain of the KI-TWPA. The idler frequency input had 40~dB attenuators at 10~mK and 3.13~K to ensure that the KI-TWPA was only presented with vacuum noise over the idler frequency range.  A second pair of diplexers at the output of the KI-TWPA routed the signal band to a MXC stage termination and the pump tone to a separate termination on a higher temperature stage to prevent the MXC from heating. The output idler band tones were passed through an isolator, a cryogenic HEMT amplifier, and a room temperature amplifier and measured using a spectrum analyzer.  The system noise in units of quanta is determined using the Y-factor relation
\begin{equation}
N_{\mathrm{sys}} = \frac{ N'_\hot(\omega_s) - Y N_\cold(\omega_s)}{Y - 1},
\end{equation}
\noindent where the $Y = P_\hot(\omega_i) / P_\cold(\omega_i)$ is the ratio of the power measured on the spectrum analyzer at the idler frequency with the hot and cold loads connected and $N'_\hot(\omega_s)= L_1 N_\hot(\omega_s) + (1-L_1)N_\term(\omega_s)$ is the effective hot load noise seen by the TWPA through the input diplexer, which has transmission coefficient $L_1 \approx 0.95$.
$N_{\hot,\cold,\term}(\omega) = \frac{1}{2}\coth(\hbar \omega / 2 k_b T_{\hot, \cold,\term})$ are the thermal noise in photon units at the temperatures $T_\hot$, $T_\cold$ and the mixing chamber temperature $T_\term$.  The contribution of the KI-TWPA to the measured noise can be extracted using a modified Y-factor
\begin{equation}
    Y' \equiv \frac{P_\hot (\omega_i) - P_{\mathrm{off}}(\omega_i)}{P_\cold (\omega_i)- P_{\mathrm{off}}(\omega_i)}
    \label{eq:yfactor},
\end{equation}
where $P_{\mathrm{off}}$ is the noise measured with the pump turned off.  The added noise contribution of the KI-TWPA is then
\begin{equation}
    A_{\mathrm{PA}} = \frac{N'_\hot (\omega_s)- Y' N_\cold(\omega_s)}{Y'-1}. 
    \label{eq:addednoise}
\end{equation}
The derivation of this equation is discussed in the appendix.
Fig.~\ref{fig:noise}.a shows system noise and KI-TWPA added noise with the pump tone at 1.65~GHz.  Fig.~\ref{fig:noise}.b shows the KI-TWPA added noise for different pump frequencies, covering the 450~MHz \textendash ~800~MHz signal band.  The added noise reaches about 1 photon, slightly higher than the quantum limit of 1/2 photon, while the system noise reaches about 2 photons. The uncertainty in the hot load temperature is estimated to be $\pm 50$mK , leading to an uncertainty of $\pm 0.25$ photons in measured system noise. The excess system noise is roughly consistent with the noise of the HEMT amplifier from the specifications, $N_\mathrm{HEMT} \sim 50$~photons for $T_N \approx 1.5$~K, and an estimate of the losses between the KI-TWPA and the HEMT. 


\begin{figure}
     \centering
     \begin{subfigure}[b]{0.9\columnwidth}
     \caption{}
         \centering
         \includegraphics[width=\columnwidth]{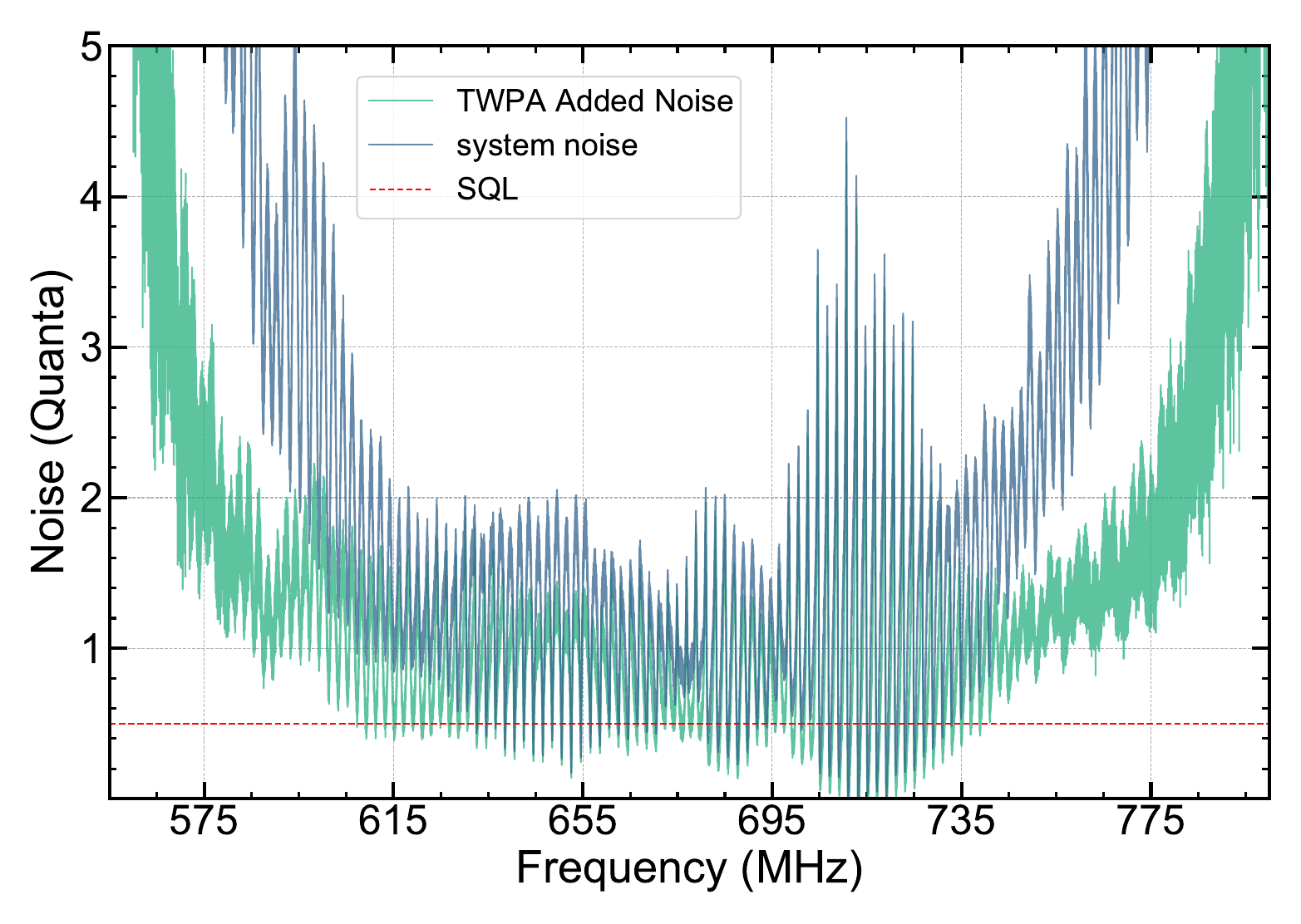}
         \label{fig:a}
     \end{subfigure}
     \begin{subfigure}[b]{0.9\columnwidth}
     \caption{}
         \centering
         \includegraphics[width=\columnwidth]{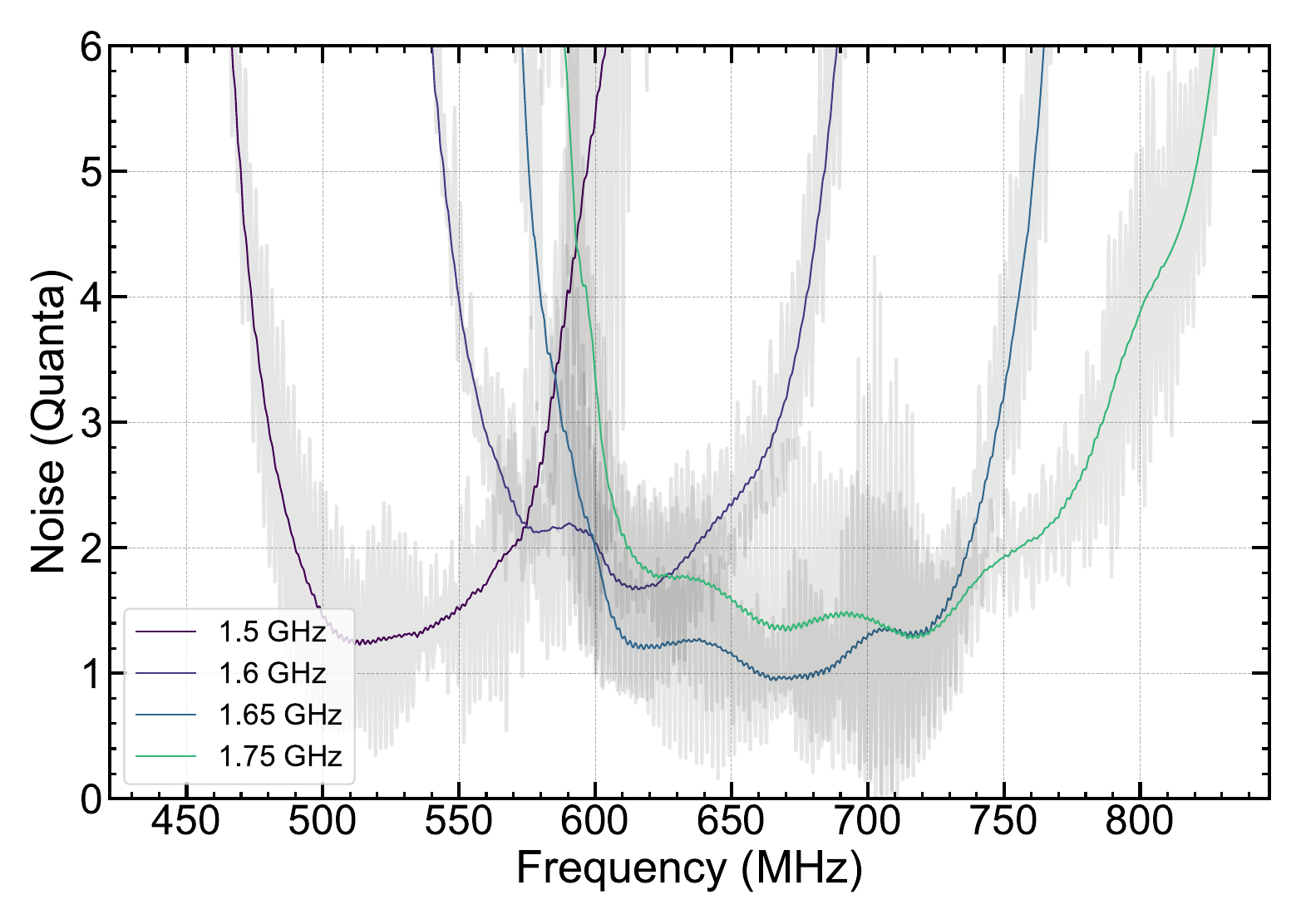}
         \label{fig:b}
     \end{subfigure}

        \caption{\label{fig:gain} a) Plot of measured total system noise and KI-TWPA added noise in units of quanta with the pump tone frequency at 1.65~GHz. The idler gain of the KI-TWPA is 18 dB for this measurement. b) Plot of the total system noise for frequencies between 500 \textendash 800~MHz. The gray is the measured data with smoothed data plotted over it. Four pump frequencies were used to cover that signal frequency range.  }
        \label{fig:noise}
\end{figure} 

\section{Conclusion}

We described the design, fabrication, and operation of a 4WM TiN Kinetic Inductance Traveling-wave Parametric Amplifier with a gain of over 20 dB that operates at frequencies below 1~GHz. The signal gain of this device is tunable as a function of pump tone frequency and covers frequencies between 450 and 800~MHz. The amplifier is most effectively operated in a frequency translating gain mode, where instead of reading out the amplified signal, the amplified idler tones were brought out of the cryostat and measured.  Operating in this manner has the important advantage of compatibility with readily available cryogenic isolators and also confers an additional power gain factor.  Using a Y-factor method, the noise was shown to be near-quantum limited between 450~MHz \textendash 800~MHz. The saturation power is approximately -70 dBm for a gain of 20~dB. The amplifier provides a solution for sub-GHz applications requiring much better noise performance than that available with transistor amplifiers.  

\begin{acknowledgments}
F.F’s research was supported by appointment to the NASA Postdoctoral Program at the Jet Propulsion Laboratory, administered by Oak Ridge Associated Universities under contract with NASA. This research was carried out at the Jet Propulsion Laboratory under a contract with the National Aeronautics and Space Administration (80NM0018D0004).  
\end{acknowledgments}

\appendix

\section{Design \& Fabrication}

\begin{figure*}

\centering
 \begin{adjustbox}{scale=0.6}
 \begin{circuitikz}[scale=0.9][american]
\ctikzset{resistors/scale=0.35,bipoles/length=1.8cm, multipoles/rotary/thickness=0.35,
ta/.style={t={\normalsize \texttt{#1}}}}
    \draw
     (0,0) rectangle (2.5,-1) node[pos=.5] {$N_\hot(\omega_s)$}
     (2.5,-0.5)--(4.5,-0.5)--(6,-0.5);
     \draw[dashed]
     (3.75,0.25)--(5.25,-1.25);
     \draw
     (4.5,-0.5)--(4.5,1.5)
     (3.25,2.5)rectangle(5.75,1.5)node[pos=.5] {$N_\term(\omega_s)$}
     (4.3,-1)node[]{$L_1(\omega_s)$}
     (0,-2) rectangle (2.5,-3) node[pos=.5] {$N_\cold(\omega_s)$}
     (2.5,-2.5)--(4.5,-2.5)--(6,-2.5)

     (7.85,-1.5)node[spdt,scale=-2.226,yscale=-1]{}
     (9.25,-1.5)--(10.5,-1.5)
     (10.8,-1.5)node[adder,scale=0.4]{}
     (10.8,-1.2)--(10.8,-0.2)
     (10.75,0.25)node[]{$A'_{PA}$}
     (11.1,-1.5)--(12,-1.5)to [amp] (14,-1.5) node[pos=1.9] {$G_{si}$}
     (14,-1.5)--(15,-1.5)--(15,-3.2)
     (15,-3.5)node[adder,scale=0.4]{}
     (15,-3.8)--(15,-5.5)--(14,-5.5)
     (12,-5.5)to [amp] (14,-5.5) node[pos=2.2] {$G_{ii}$}
     (10.5,-5.5)--(12,-5.5)
     (8,-5) rectangle(10.5,-6)node[pos=.5] {$N_\term(\omega_i)$}
     (15.3,-3.5)--(17.5,-3.5);
     \draw[dashed]
     (16.75,-2.75)--(18.25,-4.25);
     \draw
     (16.5,-3.5)--(17.5,-3.5)
     (17.5,-3.5)--(17.5,-2.5)
     (16.25,-1.5)rectangle(18.75,-2.5)node[pos=.5] {$N_\term(\omega_i)$}
     (17.3,-4)node[]{$L_2(\omega_i)$}
     (17.5,-3.5)--(19.75,-3.5)
     (20,-3.5) node[adder,scale=0.4]{}
     (20,-3.25)--(20,-2.25)
     (20,-2)node[]{$A_{HEMT}$}
     (20.25,-3.5)--(21,-3.5) to [amp](23,-3.5)node[pos=2.2] {$G_{HEMT}$} --(24,-3.5)
     
     (24.25,-3.5) node[adder,scale=0.4]{}
     (24.25,-3.25)--(24.25,-2.25)
     (24.25,-2)node[]{$A_R$}
     (24.5,-3.5)--(25.25,-3.5) to [amp](26.55,-3.5)node[pos=1.6] {$G_R$} 

      ; 
   
 \end{circuitikz}
\end{adjustbox}
 \caption{The circuit diagram of the noise model, presented in Appendix C.1, illustrates the various noise contributions in the system. $N_\hot(\omega_s)$, $N_\cold(\omega_s)$, and $N_\term$ represent the thermal noise, expressed in photon units, corresponding to the temperatures $T_\hot$, $T_\cold$, and the mixing chamber temperature $T_\term$, respectively. The term $A'_{PA}$ denotes the added noise of the TWPA, while $G_{si}$ and $G_{ii}$ represent the signal-to-idler and idler-to-idler gains, respectively. The loss between the noise sources and the device is characterized as $L_1(\omega_s)$, while the loss between the device and the HEMT amplifier is denoted as $L_2(\omega_i)$.}
 \label{fig:noisediagram}

\end{figure*}
\subsection{Device Fabrication}

A 50~nm film of TiN was sputtered on a high resistivity ($\geq$ 10,000 $\mathrm{\mu \Omega.cm}$) 150~mm silicon wafer. The TiN layer was then patterned into the microstrip layer using a 5:1 stepper photolithography method and etching in chlorine chemistry using an Inductively Coupled Plasma Reactive Ion Etcher ICP RIE. A 100~nm hydrogenated amorphous silicon layer ($\mathrm{\alpha}$-Si) was deposited on top of the microstrip layer via Plasma-Enhanced Chemical Vapor Deposition for the dielectric of the microstrip. The $\mathrm{\alpha}$-Si was then patterned and etched using an ICP RIE to define vias over the bond pads of the device. A 300~nm Nb layer was sputter deposited over the dielectric layer.
Finally, we patterned and etched the Nb "sky" plane to define the microstrip to CPW transition to the bond pads.

\section{Dielectric Loss, Extraneous Mixing Products and Gain Compression}

\begin{figure}
     \centering
     \begin{subfigure}[b]{0.9\columnwidth}
     \caption{}
         \centering
         \includegraphics[width=\columnwidth]{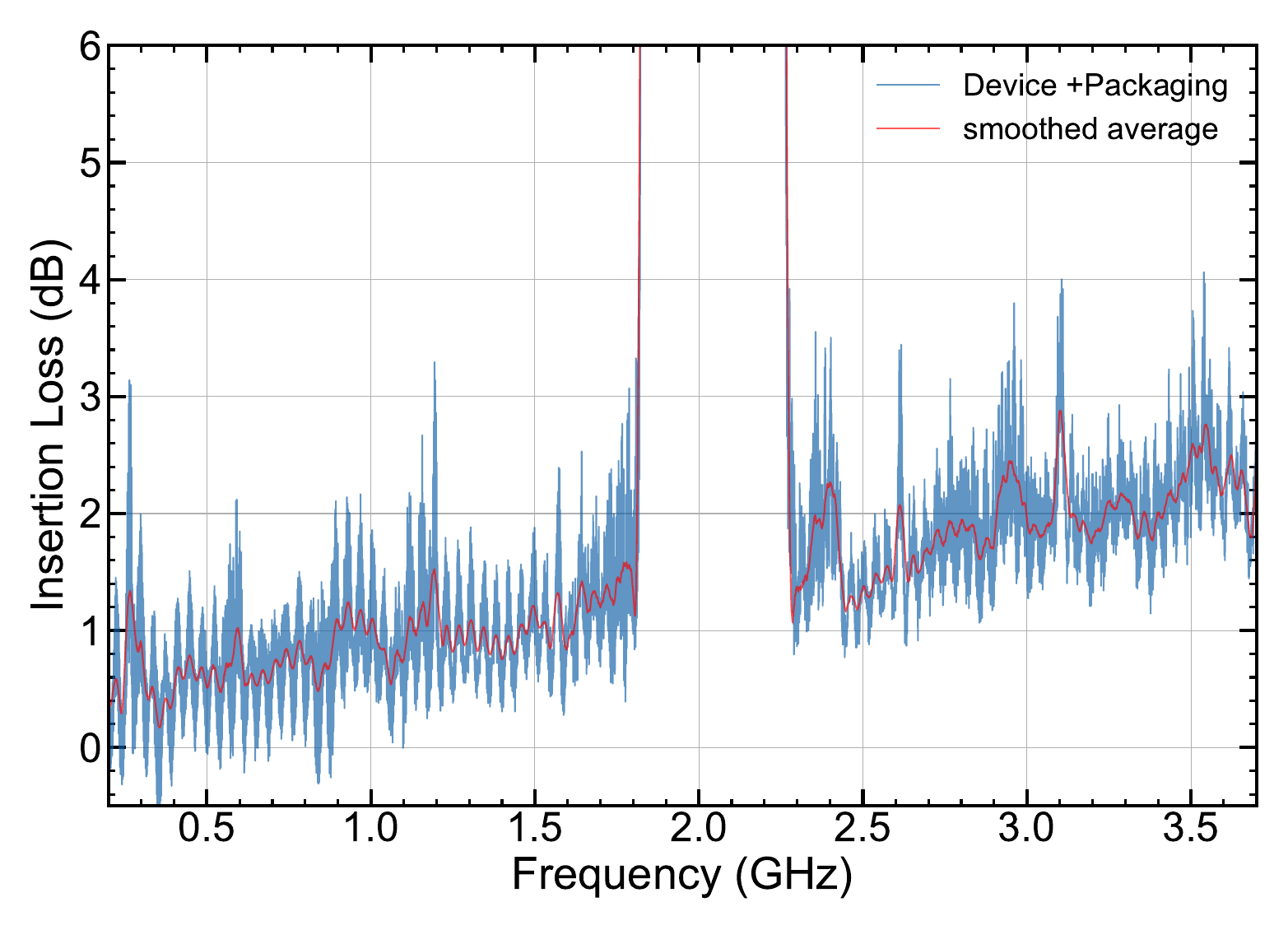}
         \label{fig:a}
     \end{subfigure}
     \begin{subfigure}[b]{0.9\columnwidth}
     \caption{}
         \centering
         \includegraphics[width=\columnwidth]{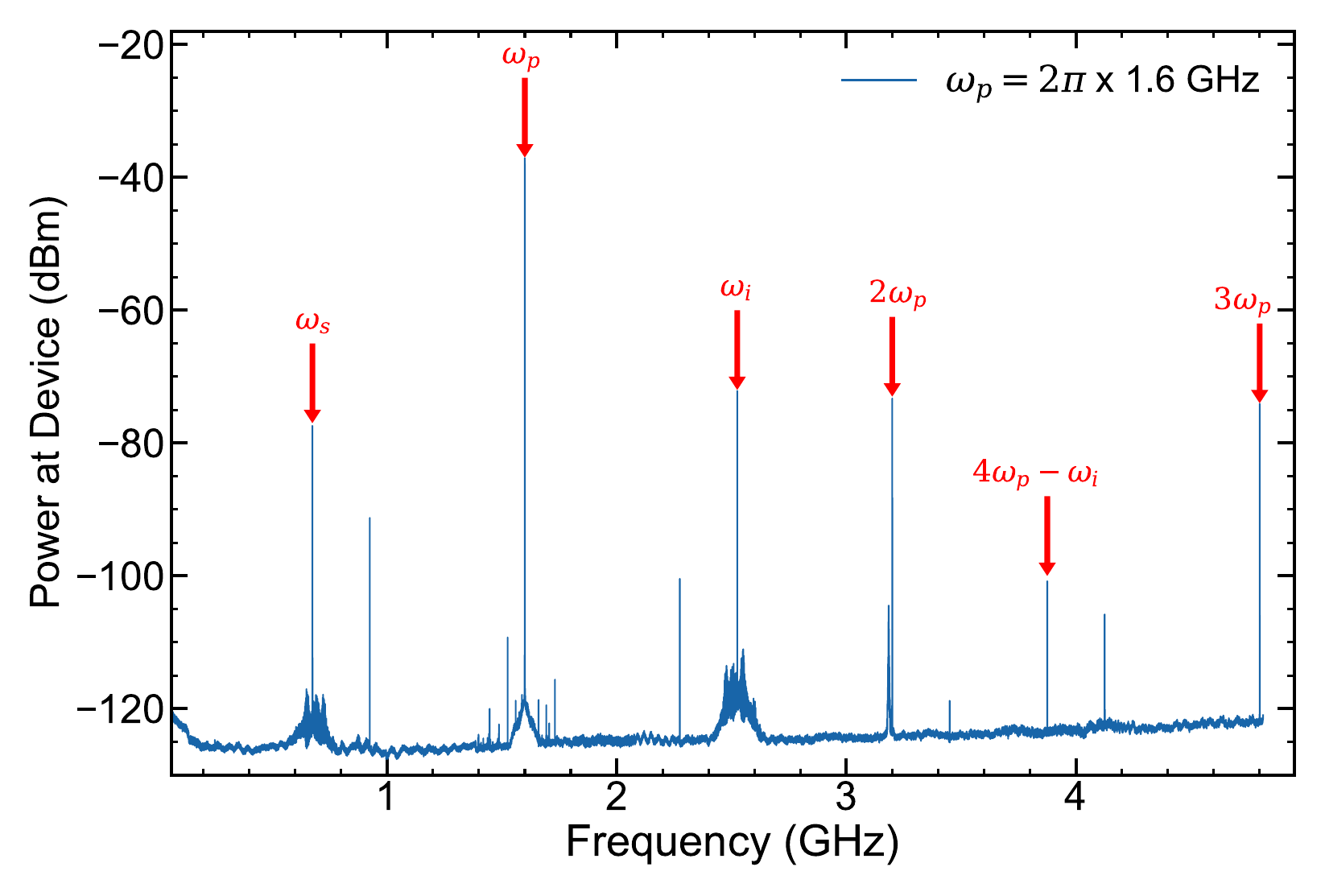}
         \label{fig:b}
     \end{subfigure}

        \caption{\label{fig:pump_off} a) Plot of measured insertion loss of the device plus the package as a function of frequency. b) Measured output power of the device at low signal power.  }
        \label{fig:noise}
\end{figure}

Fig~\ref{fig:pump_off}.a shows the measured insertion loss of the housed KI-TWPA device at a temperature of 200 mK. The loss in the signal band ranges from 0.7 to 1 dB, whereas the loss in the idler band is approximately 1.5 dB.  As single-layer resonators made from identically prepared TiN films have $Q$ factors in excess of $10^6$, it is likely that the majority of the loss is from the aSi dielectric layer, with possibly a small contribution from the normal metal transitions structures between the microwave connectors and the chip.  

Fig~\ref{fig:pump_off}.b is the measured output spectrum of the pumped device at 200~mK for a low power input signal and pump frequency of 1.6 GHz.  Higher-frequency components, including harmonics such as $3\omega_p$, $2\omega_p$, and other nonlinear mixing products, are clearly identifiable in the output spectrum of the device.  The amplitudes of the third harmonic and ``upper sideband'' tones at $4\omega_p - \omega_{s,i}$ (only $4\omega_p - \omega_i$ lies in the frequency range shown) agree well with the prediction of the coupled mode equations\cite{shibo} and the calculated dispersion.  The small amplitude of the third harmonic relative to the pump and of the upper sidebands relative to the signal, indicate that the dispersion engineering of the device is working as intended.  A stronger coupling to the upper sideband frequencies could increase the noise by mixing in vacuum fluctuations from those frequencies.  The level of the second harmonic is higher than is expected from the calculation, and the origins of the unlabeled tones are unknown.

The compression point of the amplifier was also measured for an input signal frequency of 675 MHz. The pump tone was applied at 1.6GHz, and the pump power was adjusted to attain gain levels of 10 dB, 16.5 dB, and 21.5 dB. The output power of the signal tone was measured as a function of the signal input power level. 

\begin{figure}
     \centering
         \includegraphics[width=\columnwidth]{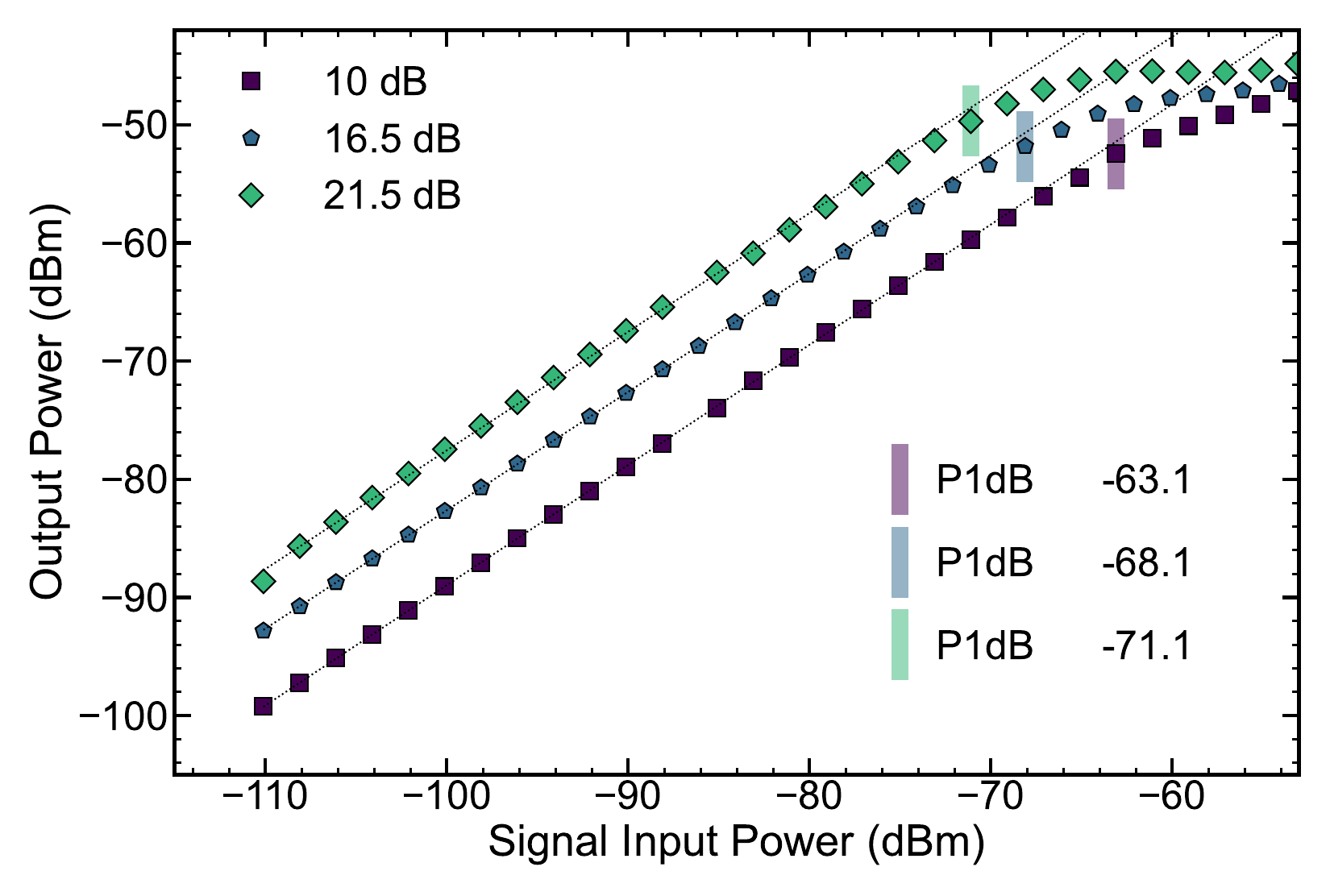}
        \caption{\label{fig:p1db} Plot of the measured output power at the device of a test tone as a function of the test tone’s input power. The measurements were performed with a pump tone at 1.6 GHz and signal gains of 10 dB, 16.5 dB, and 21.5 dB.}
        \label{fig:noise}
\end{figure}

\section{Noise Measurement setup}

\begin{figure}[htbp]
\centering
 \begin{adjustbox}{scale=0.7}
 \begin{circuitikz}[scale=0.8][american]
\ctikzset{resistors/scale=0.35,bipoles/length=1.8cm, multipoles/rotary/thickness=0.35,
ta/.style={t={\normalsize \texttt{#1}}}}
    \draw
      (-2,-3) -- (-2,-5.5)
      (-2,-3)node[ocirc]{}
      (-2,-2.5)node[]{Signal Input}
      (-2.75,-5.75) to[R] (-1.25,-5.75)
      (-2.75,-5.35)node[]{20 dB}
      (-2,-6)--(-2,-13.5)
      (-2.75,-13.75) to[R] (-1.25,-13.75)
      (-2.75,-13.35)node[]{20 dB}
      (-2.75,-14.75) to[R ] (-1.25,-14.75)
      (-2.75,-14.35)node[]{20 dB}
      (-2,-14) -- (-2,-14.5)
      (-2,-15) -- (-2,-16)

      ; \draw[blue]
      (2,-19) -- (1,-19) -- (1,-16)
      (1,-16) -- ++(-0.5,0)
      node[rotary switch <->=3 in 45 wiper -22, anchor=in, rotate = 180, color = black](R){}
      (R.out 2) -- ++(-1.1,0) 

      ; \draw
      (R.out 1) -- ++(-0.2,0) to[R] ++ (0,-1.5) node[ground]{}
     ; \draw
      (-1.5,-17.5)node[text width=1cm, align = center]{50 $\Omega$}
      
      ; \draw[blue]
      (R.out 3) -- ++(-0.2,0) -- ++(0,0.2) -- ++(1.2,0) -- ++ (0,8)
      ; \draw
      (-0.25,-5.75) to[R ] (1.25,-5.75)
      (-0.25,-5.35)node[]{20 dB}
      (0.5, -6) -- (0.5, -6.5)
      (-0.25,-6.75) to[R ] (1.25,-6.75)
      (-0.25,-6.35)node[]{30 dB}
      (0.5, -3) -- (0.5,-5.5)
      (0.5,-2)node[text width=1cm, align = center]{Hot Input}
      (0.5,-2.9)node[ocirc]{}

      (2,-19) to [twoport,ta = DPX1] (4, -19)
      (2.35,-19)node[font=\tiny,red,rotate=90]{LP}
      (1.9,-18.9)node[font=\small,red,rotate=90]{$<$ 1~GHz}
      (3,-18.5)node[font=\tiny,red]{HP}
      (3.1,-18)node[font=\small,red]{$>$ 1~GHz}
      (3.65,-19)node[font=\tiny,red,rotate=90]{C.O.M}
      (3,-17.5) -- (3,-18.23)
      
      (3,-17.5) to[twoport,ta=DPX2] (3,-15.5)
      (3,-16)node[font=\tiny,red]{LP}
      (3.1,-15.5)node[font=\small,red]{$<$ 2~GHz}
      (3.65,-16.5)node[font=\tiny,red,rotate=90]{HP}
      (4.2,-16.4)node[font=\small,red,rotate=90]{$>$ 2~GHz}
      (3,-17.1)node[font=\tiny,red]{C.O.M}
    
      (3,-14.5)--(3,-15.5)
      (3,-14.5) to [bandpass] (3,-12.5)
      (3,-12.25)node[font=\small,blue, align = center]{1 - 2~GHz}
      (3,-12.5) --(3,-7)
      (3,-6.5)--(3,-6)
      (2,-5.75)to[R](4,-5.75)
      (2.25,-5.35)node{20 dB}
      (2,-6.75)to[R](4,-6.75)
      (2.25,-6.35)node{10 dB}      
      (3,-5.5) -- (3,-4)
      (3,-0.5) node[vcoshape](SG){}
      (3,-4) to [bandpass] (3,-2) -- (SG.s)
      (SG.n) -- ++(0,1) to[R] ++ (-2,0) node[ground]{}
      

      (4, -19) to [amp, t= TWPA] (7,-19)
      (7,-19) to [twoport, ta = DPX2] (9,-19)
      (8,-18.5)node[font=\tiny,red]{LP}
      (8.6,-19)node[font=\tiny,red,rotate=90]{HP}
      (7.35,-19)node[font=\tiny,red,rotate=90]{C.O.M}
      (8,-18.25)--(8,-17)
      (8.1,-18)node[font=\small,red]{$<$ 2~GHz}
      (9.6,-18.8)node[font=\small,red]{$>$ 2~GHz}
      (9,-19)--(11,-19)
      (8,-17) to[twoport,ta=DPX1] (8,-15)
      (8,-15.4)node[font=\tiny,red]{LP}
      (8.6,-16)node[font=\tiny,red,rotate=90]{HP}
      (8,-16.6)node[font=\tiny,red]{C.O.M}
      (8.8,-16) --(9.65,-16)

      ; \draw[blue]
      (9.65,-16) -- (9.65, -10.5)
      ; \draw
      (9.65,-10.5)to[R=50$\Omega$](8.25,-10.5) node[ground]{}
      
      (9.6,-16.3)node[font=\small,red]{$>$ 1~GHz}
      (8,-15)node[font=\small,red]{$<$ 1~GHz}
      (8,-15)--(8,-13.5) to[R=50$\Omega$](7,-13.5) node[ground]{}
      
    
    (5.5,-6)--(5.5,-13.5)
    (4.5,-13.75) to[R] (6.5,-13.75)
    (4.75,-13.35)node[]{20 dB}
    (4.5,-14.75) to[R ] (6.5,-14.75)
    (4.75,-14.35)node[]{20 dB}
    (5.5,-14) -- (5.5,-16.5) -- (3.75,-16.5)

    (4.5,-5.75)to[R](6.5,-5.75)
    (4.75,-5.35)node{40 dB}
    (5.5,-5.5) -- (5.5,-3)

      (10,-19) -- (11,-19) -- (11,-18)
      (10.5,-18) rectangle (11.5,-16)
      (11,-17) node[flowarrow,rotate=90,scale=1]{}
      (12,-17) node[rotate = -90]{Isolator}
      (11,-8)--(11,-16)
      
      (11,-8) to [amp] (11,-6)
      (9.5,-7) node[]{HEMT}
      (11,-6) -- (11,-4) to [amp] (11,-1)
      (10.49,0) node[coupler2,scale=0.7, rotate=90](c){}
      (11, 1) -- (11,2)
      (10.5, 2) rectangle (11.5, 3)
      (11,2.5) node[align = center]{SA}
      (9.99,-1) to [R] (9.99,-2.5) node[tlground]{}

      (6.5, 3) rectangle (8.5, 1.5)
      (7.5,2.5) node[]{VNA}
      (7,1.75)node[ocirc]{}
      (8,1.75)node[ocirc]{}

      (3,2)node[]{Pump input}

      (8,1.75) -- (9.99, 1.75) -- (9.99,1)
      (7,1.75) -- (5.5,1.75) -- (5.5,-3)
      ;

    \draw[blue,thick,dashed] (-3.5,-12) -- (12.5,-12)
    (12.5,-12.5) node[]{MXC};

    \draw[purple,thick,dashed] (-3.5,-10) -- (12.5,-10)
    (12.5,-10.5) node[]{50 mK};    
    
    
    \draw[red,thick,dashed] (-3.5,-5) -- (12.5,-5)
    (12.5,-5.5) node[]{3.13 K};
    
     \draw
    (12.5,-2.5)node[color=brown]{300 K};
    
 \end{circuitikz}
\end{adjustbox}
 \caption{Noise measurement circuit showing the temperatures at which the different components are thermally anchored. The blue lines show the superconducting NbTi cables used to minimize loss and thermal conduction in the setup. A relay switch on the MXC was used to connect the input of the KI-TWPA to the ``Hot" and ``Cold" noise sources and the cryostat's signal input port. To measure the idler-to-idler gain, the high-pass of the DPX 2 in front of TWPA was routed out of the cryostat with 40~dB attenuation on the MXC and an additional 40~dB attenuation on the 3.13~K stage to ensure the idler input of the TWPA is effectively terminated on the MXC and the amplifier remains quantum limited.}
 \label{fig:LFsetup}

\end{figure}

\subsection{Noise Theory}

The noise power measured by the spectrum analyzer with the cryogenic switch connecting to the cold and hot terminations is
    \label{eqn:PCH}
\begin{equation}
\begin{aligned}
P_{\cold,\hot}&(\omega_i) = \\
& \bigl[ \bigl( \{ \hbar \omega_s G_{si} \left[ N'_{\cold,\hot}(\omega_s) + A'_{PA} \right] + \hbar \omega_i G_{ii} N_\term(\omega_i) \} L_2 \\ %
& + \hbar \omega_i (\{ 1 - L_2 \} N_\term(\omega_i) + A_\mathrm{HEMT}) \bigr) G_\mathrm{HEMT} + A_R \bigr] G_R B
\end{aligned}
\end{equation}
\noindent where $B$ is the bandwidth, 
$1 - L_2 \approx 0.3$ is the loss between the KI-TWPA and the HEMT amplifier and $1 - L_1 \approx 0.05$ is the loss between the KI-TWPA and the hot termination.  The small losses corresponding to $L_1$ and $L_2$ have little effect on analysis, but are included for completeness.  At 600~MHz and the operating temperature of 0.012~K, $N_T(\omega_i) = 0.5$ and $N_T(\omega_s) = 0.6$, close to the vacuum level.  $G_{si}$ is the signal to idler power gain from eqn.~\ref{eqn:Gsi}, and $G_\mathrm{HEMT}$ and $G_\mathrm{R}$ are the HEMT and room temperature amplifier gains.  The $A_\mathrm{HEMT}$ and $A_\mathrm{R}$ are the added noise of the HEMT and room temperature amplifiers in photon units.  $A'_\mathrm{PA}$ is the added noise of the KI-TWPA in excess of the noise entering at the idler frequency, as the idler noise contribution is included separately in eqn.~\ref{eqn:PCH}.  $N'_\hot = N_\hot L_1 + N_\term(\omega_s)(1 - L_1)$ is the noise from the hot load corrected for the loss $1 - L_1$, and $N'_\cold = N_\cold$ because the temperatures of the cold load and connecting lossy elements are the same.  The contributions to the total output noise are diagrammed in fig.~\ref{fig:noisediagram}.

The noise power measured with the pump off is 
\begin{equation}
    P_\mathrm{off}(\omega_i) = \hbar\omega_i \Big[\big(N_\term(\omega_i) + A_\mathrm{HEMT} \big) G_\mathrm{HEMT} + A_\mathrm{R} \Big] G_\mathrm{R}\,B.
\end{equation}
Using $G_{si} = (G_{ii} - 1)\omega_i / \omega_s$, the modified Y-factor defined in eqn.~\ref{eq:yfactor} is 
\begin{equation}
    Y' = \frac{N'_\hot(\omega_s) + A'_\mathrm{PA} + N_\term(\omega_i)} {N'_\cold(\omega_s) + A'_\mathrm{PA} + N_\term(\omega_i)}.
\end{equation}
Inverting that equation gives 
\begin{equation}
A'_{\mathrm{PA}} = \frac{N'_\hot (\omega_s)- Y' N'_\cold(\omega_s)}{Y'-1} - N_\term (\omega_i).
\end{equation}

Using the more common definition of the paramp added noise that includes the noise present at the idler frequency, $A_\mathrm{PA} = A'_\mathrm{PA} + N_\term(\omega_i)$, results in eqn.~\ref{eq:addednoise}.

\subsection{Frequency Translating Measurement Setup}

The circuit used for the frequency translating mode noise measurements is shown in Fig.~\ref{fig:LFsetup}. The signal input line has a total attenuation of 60~dB.  A hot noise source consists of two attenuators totaling 50~dB on the 3.13~K stage of the cryostat.  A coax to room temperature can be used to monitor the gain during the noise measurements.  The cold noise source is a 50~$\Omega$ termination on the mixing chamber stage. All three of those inputs are connected to a switch with the common port connecting to a 1~GHz diplexer (TTE, P/N: D1G09G01)~\cite{TTE}.  Another diplexer (DPX2) (TTE, P/N: D02G18G2) is used to inject the pump tone while routing the idler frequencies to 40~dB of attenuation on the mixing chamber stage, which serves as the idler input termination of the KI-TWPA.  Those attenuators also connect to a cable at room temperature that can be used to measure the idler-to-idler frequency gain. The pump tones were generated using a National Instruments QuickSyn microwave synthesizer (Model: FSW-0020). A cold fixed frequency bandpass filter (Fairview, P/N: FMFL1012) with a center frequency of 1.5~GHz and bandwidth of 1~GHz and a room temperature variable frequency bandpass filter are used on the pump line to further limit any extra noise in the signal and idler bands from the synthesizer in excess of 300~K noise.

 Another set of the same diplexers is used at the output of the KI-TWPA to terminate the signal and pump tones at the mixing chamber and 50~mK stages, respectively, and to route the idler band through an isolator (PAMTECH, P/N: 55387, CTB1299KG) and a cold HEMT amplifier and out of the cryostat. The two sets of two diplexers could easily be combined into two triplexers.  A room temperature low noise amplifier (Minicircuits, P/N: ZVA-183-S+) was used to further amplify the idler frequencies. Using a 10~dB directional coupler, the idler frequency was routed to a VNA for gain measurements and sent to a spectrum analyzer for noise measurements.

\hspace{-1cm}
\begin{figure}

\centering
 \begin{adjustbox}{scale=0.6}
 \begin{circuitikz}[scale=0.9][american]
\ctikzset{resistors/scale=0.35,bipoles/length=1.8cm, multipoles/rotary/thickness=0.35,
ta/.style={t={\normalsize \texttt{#1}}}}
    \draw
     (0,0) node[vcoshape](SG){}
     (0,-0.73)--(0,-2)--(0.5,-3)--(1.5,-3)
     (0,-2)--(-0.5,-3)--(-0.5,-5)node[ocirc]{} 
     (1.5,-3)--(2.5,-3) 
     (3.2,-3) node[mixer]{}
     (3.2,-2.3)--(3.2,-0.73)
     (3.2,0) node[vcoshape](SG){}
     (3.2,-3.7)--(3.2,-5)node[ocirc]{}
     (3.9,-3)--(7,-3)
     (7.7,-3) node[mixer,boxed](m){}
     (7.7,-3.7)--(7.7,-5)node[ocirc]{}
     (7.4,-2.3)--(7.4,-1)
     (8,-2.3)--(8,-1)
     (0,1)node[]{$\omega_p$}
     (-0.5,-5.5)node[]{Pump input}
     (3.2,-5.5)node[]{Input drive}
     (3.2,-5.8)node[]{$\omega_s$}
     (3.2,1)node[]{$\omega_p - \omega_s$}
     (2.1,-2.7)node[]{LO}
     (4.7,-2.7)node[]{USB($\omega_i$)}
     (2.6,-4)node[]{LSB}
     (7.7,-5.5)node[]{Cryostat output}
     (7.7,-5.8)node[]{$\omega_i$}
     (9.4,-3)node[]{IQ Mixer}
     (8,-0.5)node[]{Q}
     (7.3,-0.5)node[]{I}

     ;

 \end{circuitikz}
\end{adjustbox}
 \caption{Circuit diagram for homodyne measurements of the phase shift of a resonator utilizing the KI-TWPA in frequency translating mode.}
 \label{fig:ftgreadout}

\end{figure}

While not implemented in the work reported here, the typical circuit used, for example, for homodyne measurements of the phase shift of a resonator, can be easily adapted for frequency translating operation.
A possible circuit is shown in Fig.\ref{fig:ftgreadout}.  Instead of generating the resonator probe tone directly, a synthesizer generates a tone at the difference frequency $\omega_p - \omega_s$.  That tone is mixed with the pump frequency to generate the $\omega_s$ probe as the lower sideband (LSB) and the idler tone $\omega_i$ as the upper side band (USB).  The sideband separating mixer could be implemented in various ways including with two 90 degree hybrid couplers and two standard mixers or with a single mixer and two band pass filters.  The USB is then used as the LO for the conventional quadrature mixer at the output of the cryostat.


\bibliography{apssamp}

\end{document}